\documentclass[conference]{IEEEtran}
\IEEEoverridecommandlockouts
\usepackage{cite}
\usepackage{amsmath,amssymb,amsfonts}
\usepackage{algorithmic}
\usepackage{graphicx}
\usepackage{textcomp}
\usepackage{xcolor}
\usepackage{subcaption}
\usepackage{gensymb}
\usepackage[section]{placeins}

\usepackage{xcolor}
\usepackage{flushend}
\def\BibTeX{{\rm B\kern-.05em{\sc i\kern-.025em b}\kern-.08em
		T\kern-.1667em\lower.7ex\hbox{E}\kern-.125emX}}
\begin{document}
	
	\title{Performance Analysis of Fixed Broadband Wireless Access in mmWave Band in 5G}
	
	\author{\IEEEauthorblockN{Soumya Banerjee}
		\IEEEauthorblockA{\textit{Virginia Modeling Analysis and}\\
			\textit{Simulation Center} \\
			\textit{Old Dominion University}\\
			Norfolk, VA, USA \\
			s1banerj@odu.edu}\\ % 0000-0001-6455-5134
		\and
		\IEEEauthorblockN{Sarada Prasad Gochhayat}
		\IEEEauthorblockA{\textit{Department of Electrical and}\\
			\textit{Computer Engineering,} \\
			{Villanova University,}\\
			Villanova, PA, USA \\
			sgochhay@villanova.edu }
		\and
		\IEEEauthorblockN{Sachin Shetty}
		\IEEEauthorblockA{\textit{Virginia Modeling Analysis and}\\
			\textit{Simulation Center} \\
			\textit{Old Dominion University}\\
			Norfolk, VA, USA \\
			sshetty@odu.edu}% 0000-0002-8789-0610
	}
	
	\maketitle
	
	\begin{abstract}
		An end-to-end fiber-based network has the potential to offer multi-gigabit fixed access to end-users. However, a fiber network’s biggest hurdle is delivering that fiber access to the end-user. Especially in places where fiber is non-existent, it can be time-consuming and costly to deploy, resulting in Operators experiencing a long delay in realizing a return on their investment. 
		
		This work investigates transmission data from fixed broadband wireless Access in mmWave band in 5G. With the increasing interest in this domain, it is worth investigating the transmission characteristic of the data and utilizing the same to build more sophisticated capabilities. Existing datasets for mmWave band are detailed but generated from simulated environments.
		In this work, we introduce a dataset built from the collection of real-world transmission data from Fixed Broadband Wireless Access in mmWave Band device(RWM6050). The goal of this data is to enable self-configuration capability based on transmission characteristics. Towards this goal, we present an online machine learning-based approach that can classify transmission characteristics with real-time training.
		We also present two more advanced temporal models for more accurate classifications. 
		We demonstrate that it is possible to detect the transmission angle and distance directly from the analysis of transmission data with very high accuracy. We achieved up to $99\%$  accuracy on the combined classification task.
		Finally, we outline some interesting future research scopes based on the collected data.
		
	\end{abstract}
	
	\section{Introduction}
	
	As operators continued to trial various technologies in recent years, it has become clear that initial 5G deployments will begin with fixed broadband wireless access in the mmWave band. This is partly driven by a large amount of spectrum available in both licensed and unlicensed bands and the maturity of technologies that facilitate mmWave solutions. In addition, the expected increase in capacity and the enhancement in QoS resulting from this situation will benefit the operators as this will generate additional revenue from new business opportunities built around various 5G use cases, such as virtual reality, real-time distributed UHD gaming, and tactile internet for remote surgery.
	
	As operators continued to trial various technologies in recent years, it has become clear that fixed broadband wireless access (which tries to replace the end-to-end fiber-based connection) in the mmWave band \cite{bae2014architecture} will be an essential part of the initial 5G deployment\cite{sakaguchi2017and}. This is partly driven by the availability of a large amount of spectrum in both licensed and unlicensed bands. Furthermore, the advancement of technologies that foster mmWave solutions paves the way for fixed broadband wireless access. In addition, the expected increase in capacity and the enhancement in QoS resulting from this situation will benefit the operators as this will generate additional revenue from new 5G business opportunities, such as real-time gaming, virtual reality, and tactile internet for remote surgery.

	\begin{center}
		\begin{figure}[h]
			\includegraphics[width=\linewidth]{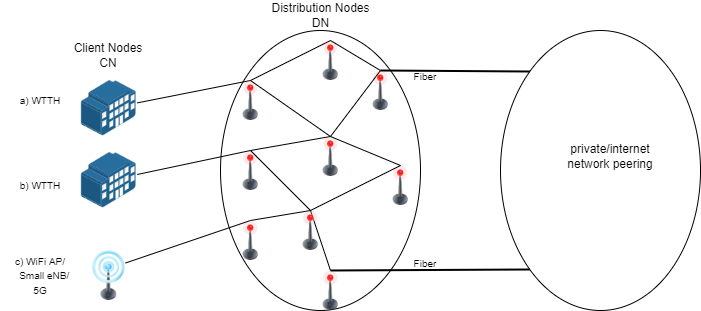}
			\caption{Fixed wireless broadband access}
			\label{1.1}
		\end{figure}
	\end{center}
	
	\subsection{Fixed Wireless Broadband Access}
	Access to spectrum is an important first step to enabling wireless technology as a solution to fixed access networks. The second “wave” or step is the significant technological advancement in antenna design and wireless communication protocols based on beamforming, MIMO, and phased array targeted to mmWave channels. And lastly, fueled by an industry-driven ecosystem and standardization bodies such as WiFi Certified WiGig, IEEE 802.11, ETSI ISG mWT, and 3GPP, there have been considerable improvements in signal processing techniques that opened opportunities to create mmWave solutions that are commercially feasible.
	
	Operators today offer fixed wireless access primarily using fiber-DSL, cable, wireless, and satellite mediums with a limited data rate between a few megabit/sec to hundreds of Mbps, depending on the location and the access technology in use. For example, xDSL-based fixed access technology offers a data rate of hundreds of Mbps. However, the services are geographically challenged and cannot be guaranteed uniformly over distance in urban and suburban areas with the required quality of service or throughput. On the other hand, satellite-based access technology can cover remote areas and reach 100s of Mbps based on the DVB-S2x specification. However, due to the inherent delay in GEO-based Satellite networks, this type of fixed access service will always have limitations.

	Among the available technologies, an end-to-end fiber-based network has the potential to offer multi-gigabit fixed access to end-users. However, a fiber network's biggest hurdle is delivering that fiber access to the end-user. In communities where fiber is non-existent, it can be time-consuming and costly to deploy, resulting in Operators experiencing a long delay in realizing a return on their investment. In most cases, the fiber is never seen at the end-user location. However, the fiber-based link is terminated at a central hub, and the last mile access is provided by either xDSL, copper, or wireless technologies limiting the last mile throughput significantly. Figure \ref{1.1} shows the deployment setting of the multi-gigabit fixed access network.
	
\begin{center}
		\begin{figure}[h]
			\includegraphics[width=\linewidth]{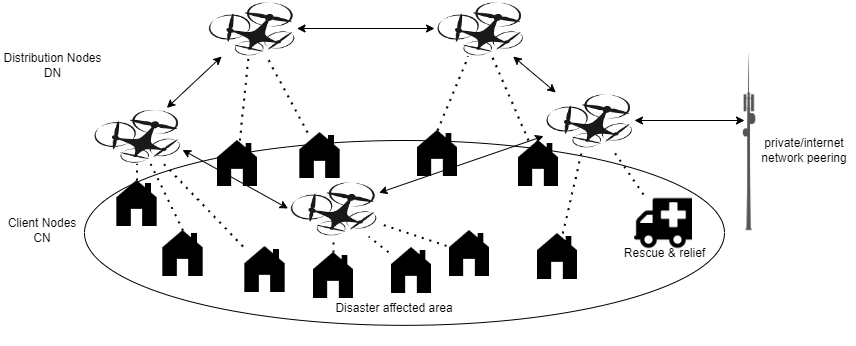}
			\caption{mmWave network in a disaster relief scenario}
			\label{1.2}
		\end{figure}
	\end{center}

	We envision that this technology will be helpful and can be adapted for mobile deployment and will help provide better network capabilities in areas with poor or uncertain connections.
	For example, when soldiers are deployed and need to communicate in adverse terrains the stability of network resources becomes very important. mmWave devices deployed on UAVs can guarantee high bandwidth network access.
	Similarly in disaster relief scenarios, like after earthquakes and tsunamis, when the telecommunication infrastructure has been destroyed, the relief effort can be coordinated over a similar mmWave-UAV network. Figure \ref{1.2} shows the deployment of a drone-based mmWave network in a disaster relief scenario.
	
	Wireless technologies and spectrum allocation have undergone radical changes in the last several years. In particular, the availability of mmWave spectrum from 26 to 90GHz has opened a significant opportunity to deliver multi-gigabit wireless access to users. For example, the FCC recently opened up around 3.85GHz of the licensed spectrum between the 27.5 to 40GHz band (27.5–28.35GHz, 37–38.6GHz, 38.6–40GHz). Furthermore, 7GHz of spectrum in the unlicensed V-band (64–71 GHz) was added to the existing unlicensed V-band (57–66GHz) spectrum, and there exists 10GHz of spectrum in the lightly licensed E-band (71–76GHz and 81–86GHz). This uniquely provides nearly 28GHz of spectrum for use in wireless technologies.

	For the realization of Fixed wireless Broadband access, two types of nodes are needed, namely the access node and aggregation node. The access nodes could be installed on the rooftop, near the window, or inside the home. Similarly, the aggregation nodes could be installed on the street-side lamp post. In addition, these aggregation nodes could be connected with mesh topology to support redundancy and mitigate interference. It's desirable that such mesh networks include self-configuration capabilities because their presumed deployment scenarios will be dynamic. Building such capabilities is non-trivial and requires investigation. Specifically, for self-configuration capabilities, the transmitting-receiving radio pair need spatial awareness. It is always conceivable to include additional hardware to address the issue, but doing so for each transceiver might quickly become economically untenable. A possible solution is to infer this information from the radio transmission itself.

	\begin{figure}
		\centering
		\includegraphics[width=0.35\textwidth]{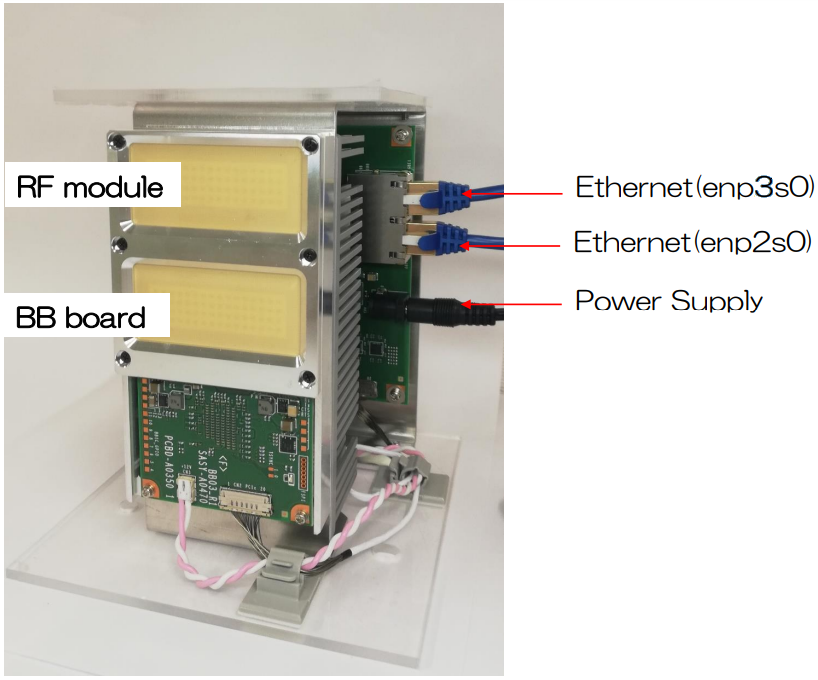}
		\caption{RWM6050 device}
		\label{1.3}
	\end{figure}

	Existing mmWave datasets are either from a user case where it is being utilized as a radar \cite{singh2019radhar, wei20213dried, kramer2022coloradar} or were generated using simulations \cite{alkhateeb2019deepmimo, agrawal20195th}. While simulation data are invaluable for research, they cannot substitute the need for actual physical experiments.  
	In this paper, we discuss data collection Fixed Broadband Wireless Access in mmWave Band device, i.e., RWM6050. Especially, we collected the statistical data between transmitter and receiver at various distances and angles of incidence between the devices. Subsequently, we investigated machine learning approaches to classify various transmission distances and angles of incidents. We publicly release \footnote{https://github.com/soumyaxyz/} the data and codes for investigation by the community.% and dynamically self-configure to obtain a higher data rate.  

	\begin{figure}[h]
		\includegraphics[width=\linewidth]{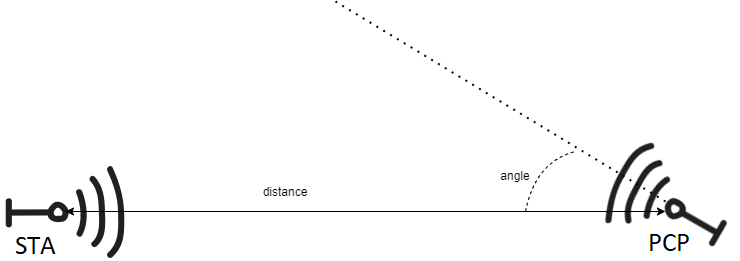}
		\caption{The data collection setup}\label{2.1}
	\end{figure}
	
	The remainder of the paper is organized into three sections. The following section describes the experimental setup for the data collection. In the subsequent section, the machine learning-based analysis of the dataset is presented. Finally, the conclusion summarizes the work and outlines future research opportunities based on this work.

	\section{Experiment Setup}
	In this section, we describe the experimental setup for data collection using the RWM6050 device (See figure \ref{1.3}). 
	
	\begin{figure}[htb!]
		\centering
		\begin{subfigure}[t]{0.45\linewidth}
			\centering
			\includegraphics[width=\linewidth]{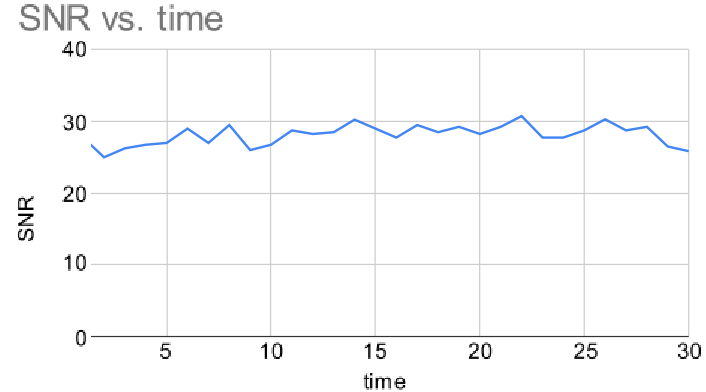}
			\caption{SNR vs Time}
			\label{3.a}
		\end{subfigure}%
		~ 
		\begin{subfigure}[t]{0.45\linewidth}
			\centering
			\includegraphics[width=\linewidth]{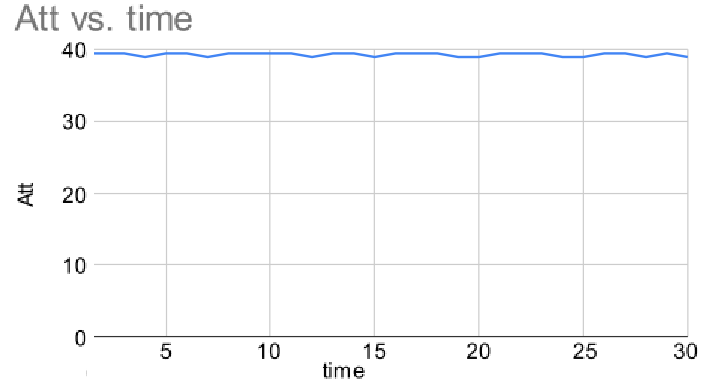}
			\caption{Attenuation Vs Time}
			\label{3.b}
			
		\end{subfigure}
		\caption{Statistic of gathered data}
	\end{figure}
	
	\subsection{Hardware Description}
	The RWM6050 software supports the transfer of TCP/IP data packets between RWM6050-based modules. This device has two components namely: PBSS Control Point (PCP) and a station (STA). By configuring the RWM6050 device it is possible to establish the point-to-point and point-to-multipoint wireless links. RWM6050 runs on Linux operating system. Two important Linux tools, namely hostapd and wpa-supplicant, are used to establish links. Hostapd is a Linux daemon that configures a node as a PBSS Control Point (PCP). Similarly, the wpa-supplicant configures a node as a station (STA).  We need to configure IP addresses for FMWS-module1 (PCP) and FMWS-module2 (STA) in the same network.
	By configuring the RWM6050 device, we established the point-to-point and point-to-multipoint wireless connection. The RWM6050 firmware supports the transfer of TCP/IP data packets between RWM6050-based modules, where the PCP is a node as a PBSS Control Point and associates the other node to the PCP as a station (STA). %In addition, in the mechanism of the RWM6050, there is a  user space that controls the RWM6050 and performs operations such as setting the operational mode, scanning, and joining a BSS. User agents (such as iw, wpa-supplicant, and hostapd) links.  %We utilized the SiBeam Sil6342 type radio. 

	\subsection{Experimental Design}
	We envisioned that the deployment of mmWave devices in a real-world setting will invariably require transmission through noisy environments. We were limited to an indoor setting as our setup needed to be tethered for power and connections. We deployed the devices in the atrium of our institute. The setting provided a mostly unobstructed open environment. The typical personnel traffic parallels the expected noise in a real-world deployment. 
	
	\begin{figure}[ht]
		\includegraphics[width=\linewidth]{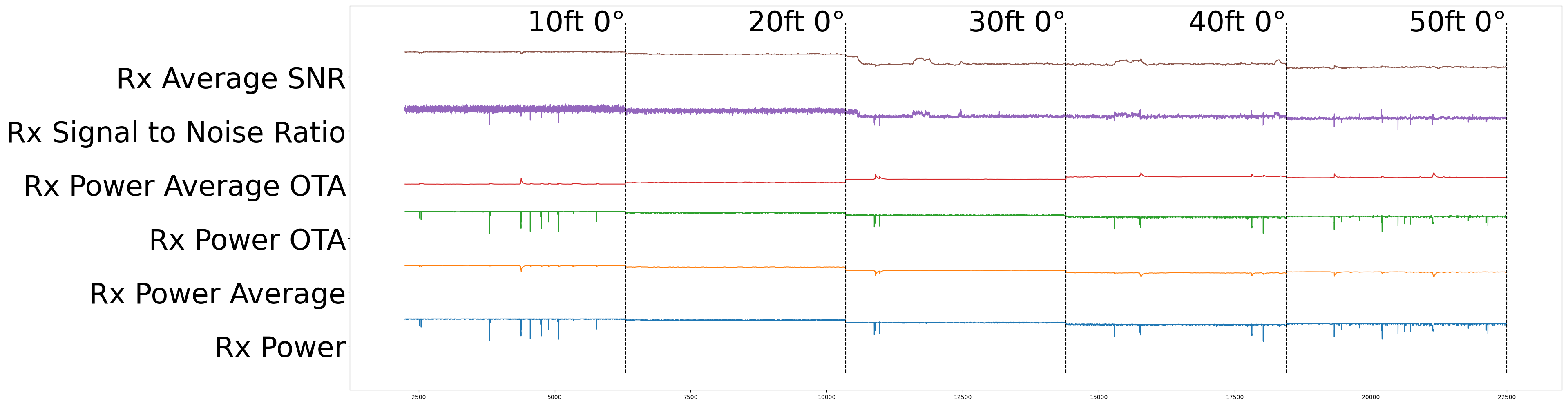}
		\caption{Feature variation w.r.t transmission distance}
		\label{3.1}
	\end{figure}
	
	\begin{figure}[ht]
		\includegraphics[width=\linewidth]{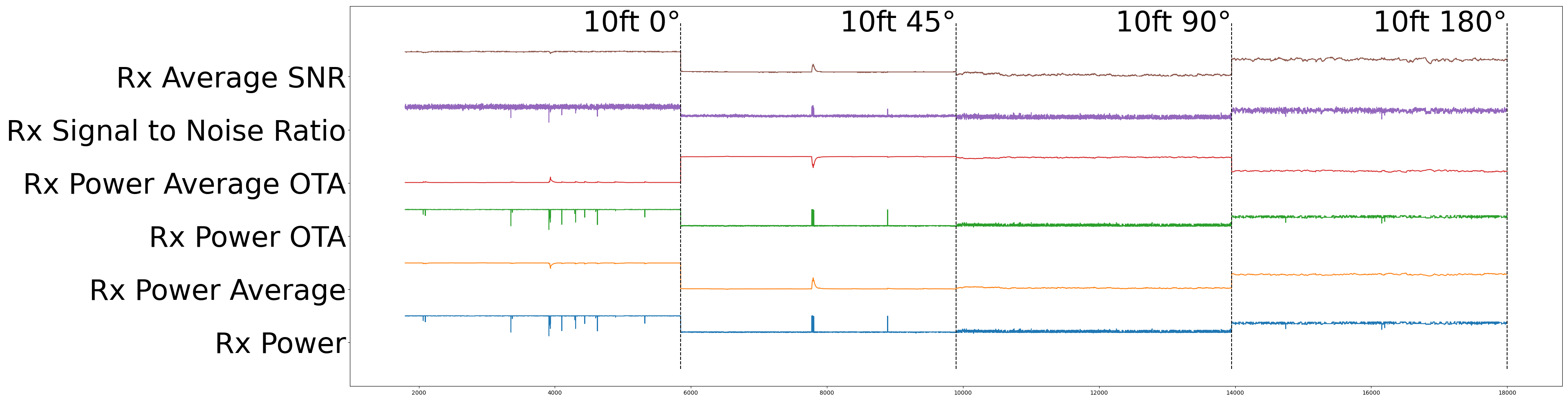}
		\caption{Feature variation w.r.t transmission angle}
		\label{3.2}
	\end{figure}
	
	\subsection{Data Aggregation}
	For data aggregation, the  PCP and STA are $10$ to $50$ feet apart on a platform with different angles. We connected with ssh over Ethernet(enp3s0) to both PCP and STA and sent messages from PCP to STA and vice versa. We sent ICMP, TCP, and UDP packets using ping and iperf.

	\begin{figure}[!htb]
		\centering
		\includegraphics[width=.7\linewidth,height=.35\linewidth]{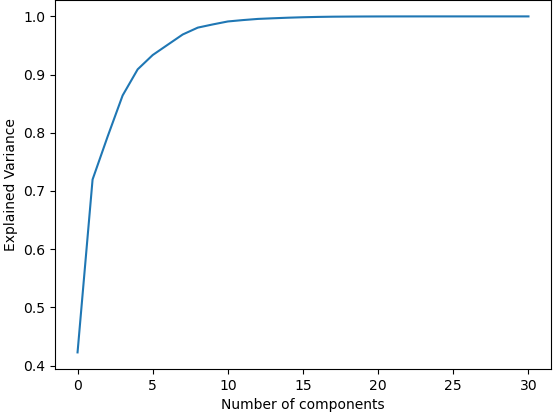}
		\caption{PCA component-wise variance}\label{3.3}
	\end{figure}
	
	RWM6050 software supports the retrieval of statistics on link performance, which is very helpful for getting the average received power, average Signal-to-Noise Ratio (SNR), and average Packet Error Rate (PER).  We have shown the SNR and attenuation data in Figure \ref{3.a} and \ref{3.b}.  For training the machine learning model, we collected transmission data between the PCP and STA nodes in $20$ different settings. $10$, $20$, $30$, $40$, and $50$ feet distance between the radios and $0$\textdegree, $45$\textdegree, $90$\textdegree, and $180$\textdegree  angle subtended by the transmitter as shown in figure \ref{2.1} for each of the transmission distances.

	\section{Identification of transmission configuration based on machine learning}
	In this section, we present our investigations into the identification of transmission configurations. First, we present our investigation into feature selection and feature extraction to filter out the extraneous features in the dataset. 
	Then, we present a study on online classification of the transmission data, suitable for real-time training. Finally, we present our study of offline iterative classification on which we achieve the highest accuracy.

	\begin{figure*}[!htb]
	\includegraphics[width=\linewidth]{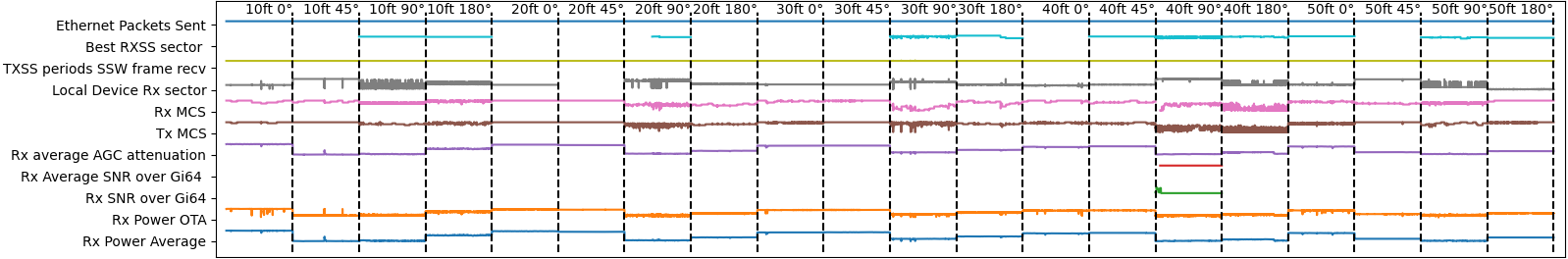}
	\caption{Feature variation in MRMR features}\label{3.4}
	\end{figure*}

	\subsection{Feature selection and feature extraction}
	Analyzing the data collected from the aforementioned experiments, it is apparent that the observed features vary proportionally with the mode of transmission. Of the $31$ features collected according to our empirical observation the most significant information should be encoded in the following features: Rx Power, Rx Power Average, Rx Power OTA, Rx Power Average OTA, Rx Signal to Noise Ratio, Rx Average SNR. Figures \ref{3.1} and \ref{3.2} show how the features vary when the transmission distance and angle are varied.
	
	We performed PCA (Principal component analysis) on the training data and, as demonstrated in figure \ref{3.3}, $10$ components yields the most optimal values. Data transformed with a $10$ component PCA yields The best results.

	MRMR (Maximum Relevance, Minimum Redundancy) \cite{ding2005minimum} is a widely adopted feature selection algorithm. When we execute MRMR on the data it identifies the following $10$ as the most significant features:
	Tx MCS, Best RXSS sector, Rx average AGC attenuation, Ethernet Packets Sent, Rx SNR over Gi64, Local Device Rx sector, Rx Power OTA, Rx Power Average, Rx Average SNR over Gi64, TXSS periods SSW frame recv.
	Figure \ref{3.4} demonstrated the feature variation in MRMR features. For clarity, where the features assume a zero value, it's not plotted. We can see that MRMR selects several features that are very discriminative for some of the classes but are uniformly zero in most.  In one of our experiments, we observed that the MRMR feature performs better during training but performs poorer than the empirically selected features during testing.

	We collected $5000$ data points for each of the $20$ classes. We additionally collected $1000$ data points from $25$ and $35$ feet distances. Of the $31$  features collected, some features are counter that count up over the transmission lifetime and thus are not suitable for training machine learning models. We normalized those features by taking the difference between the successive value as the final value.

	\subsection{Online classification}
\begin{figure}
		\includegraphics[width=\linewidth]{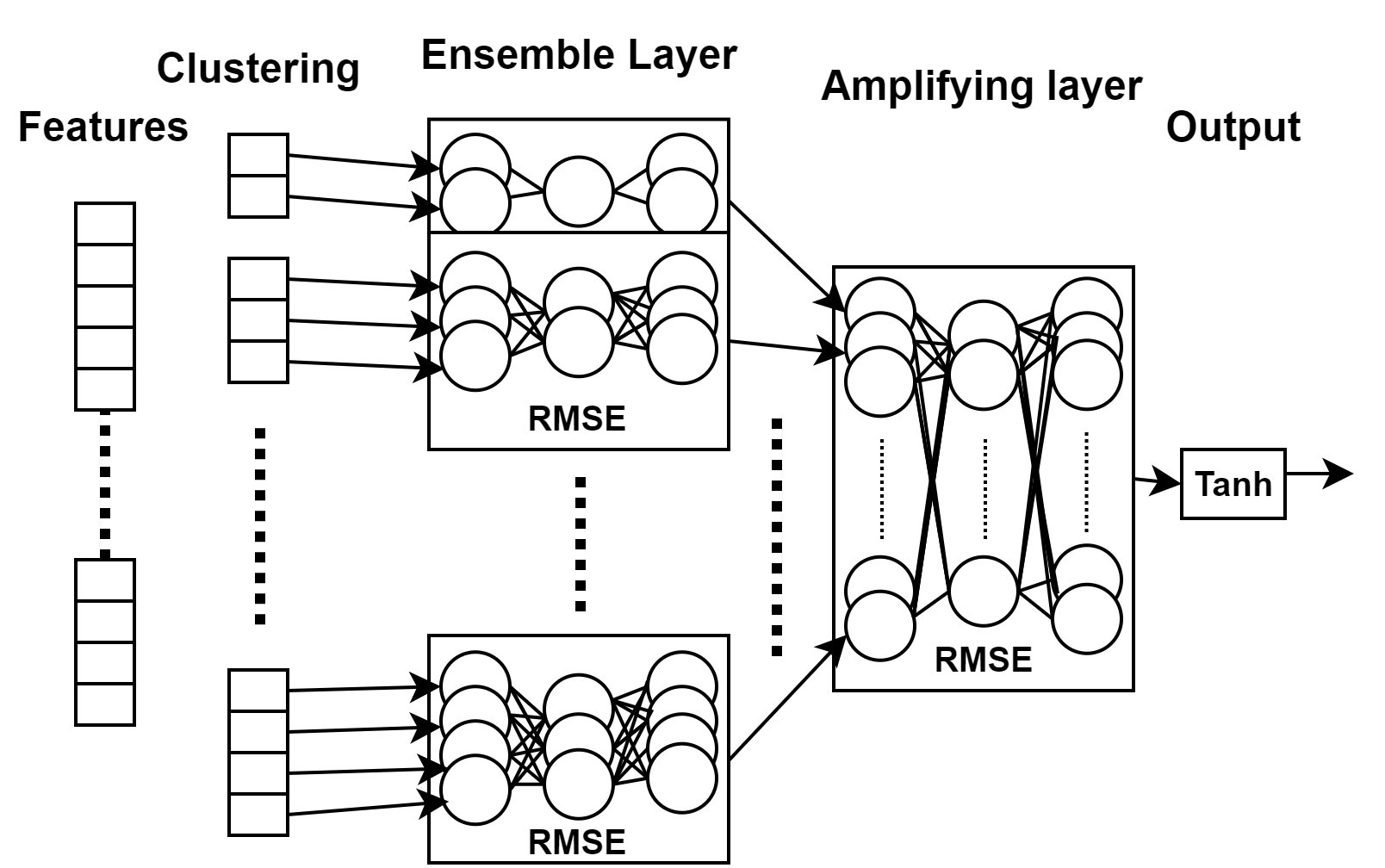}
		\caption{The Kitsune architecture}\label{7}
	\end{figure}

	In this section, we perform an online classification of the data with the caveat that the data must be processed in a single pass. We applied Kitsune, a lightweight online machine learning architecture proposed by Mirsky \textit{et al.} \cite{mirsky2018kitsune}. The Kitsune architecture utilizes a hierarchy of auto-encoders trained on in-distribution data. The out-of-distribution data can be identified by a high reconstruction error. Figure \ref{7} shows a schematic diagram of Kitsune.

	% \begin{figure}[ht]
		% \includegraphics[width=\textwidth]{ensemble-autoencoder.png}
		% \caption{The Kitsune architecture}
		% \label{3.4}
		% \end{figure}

\begin{figure}[ht]
		\includegraphics[width=\linewidth]{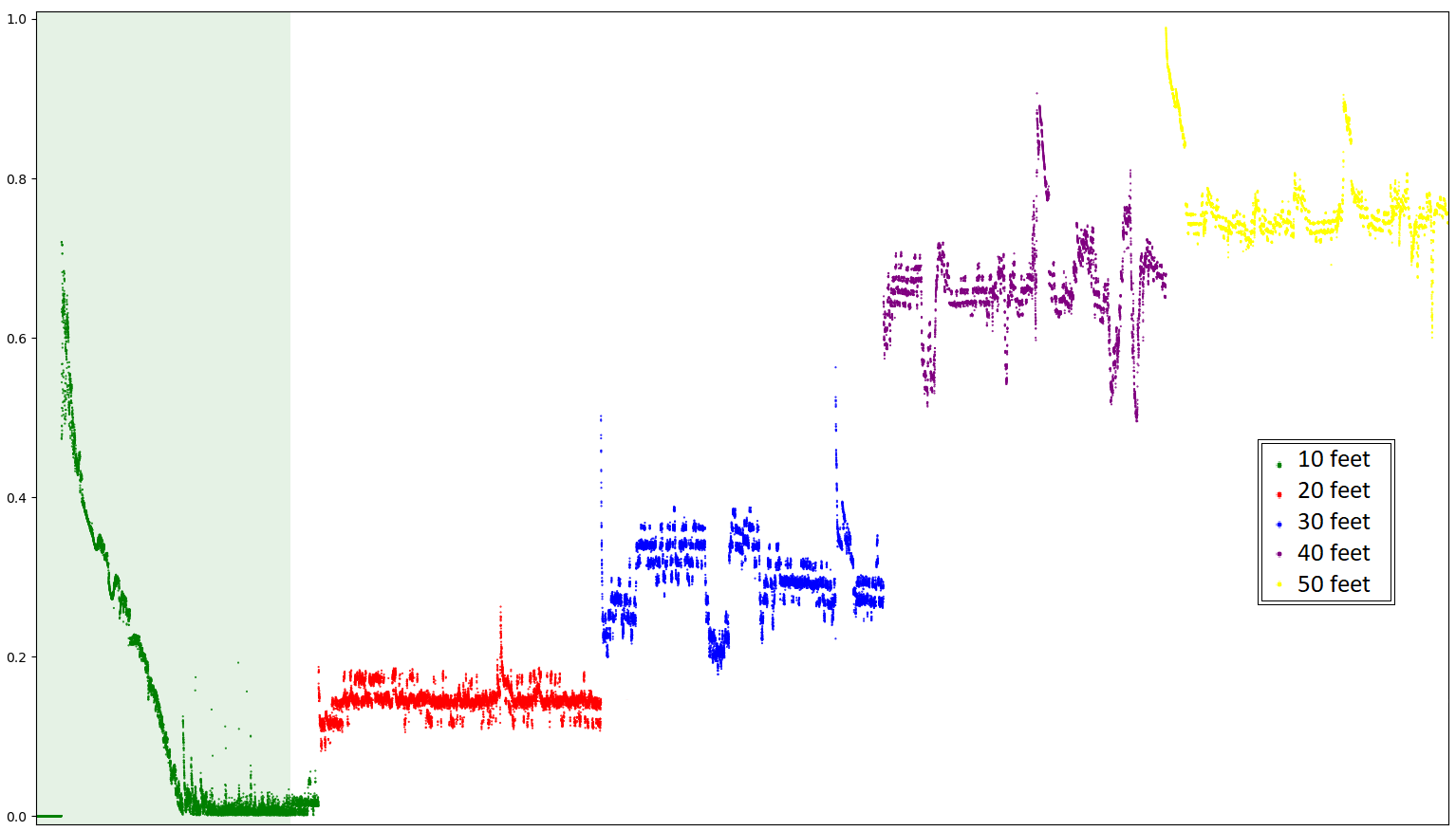}
		\caption{linear relationship between prediction and the transmission distance}
		\label{8}
	\end{figure}

	We trained the model with only data of 10 feet transmission distance. Then evaluated the remaining distances. The reconstruction error demonstrated a linear relationship to the transmission distance. Figure \ref{8} shows the results. However, a similar trend is not observed for transmission angles.

	% \floatbarrier

\begin{figure}
		\begin{subfigure}[h]{0.49\linewidth}
			\includegraphics[width=\linewidth]{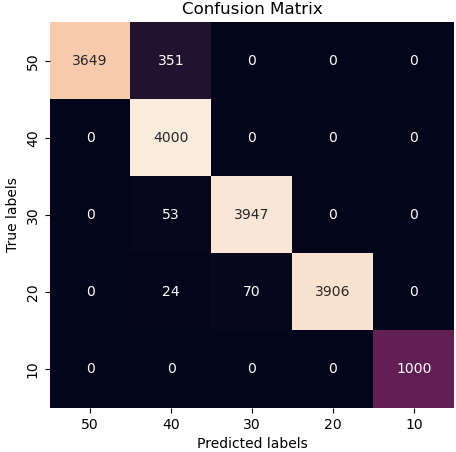}
			\caption{Full training data}
		\end{subfigure}
		\hfill
		\begin{subfigure}[h]{0.49\linewidth}
			\includegraphics[width=\linewidth]{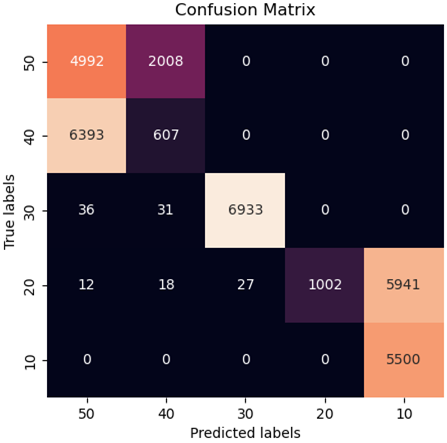}
			\caption{Half training data}
		\end{subfigure}
		\caption{Predicting transmission distance}
		\label{3.6}
	\end{figure}

	To address the issue we trained an ensemble of Kitsune where each sub-network is trained on one class of data. For the case where only communication distance is considered, the model achieved an accuracy of 0.97. If we decrease the training data by half the accuracy falls to 0.568. However, from the confusion matrix, figure \ref{3.6}a and b, we see that the inter-class misclassification is localized to the more extreme cases. 
	% 10> 0.946
	% 20> 1
	% 30> 0.646
	% 40> 0.504
	% 50> 1

\begin{figure}
		\begin{subfigure}[h]{0.32\linewidth}
			\includegraphics[width=\linewidth]{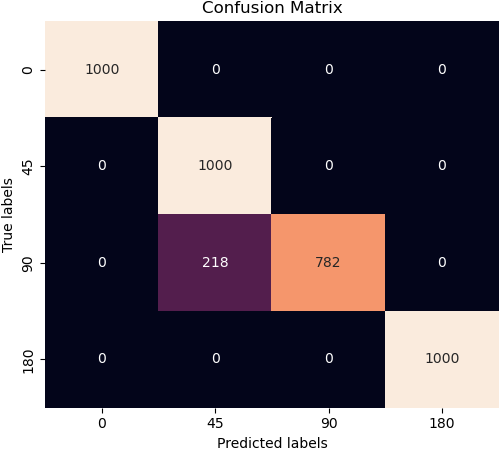}
			\caption{Distance 10ft}
		\end{subfigure}
		\begin{subfigure}[h]{0.32\linewidth}
			\includegraphics[width=\linewidth]{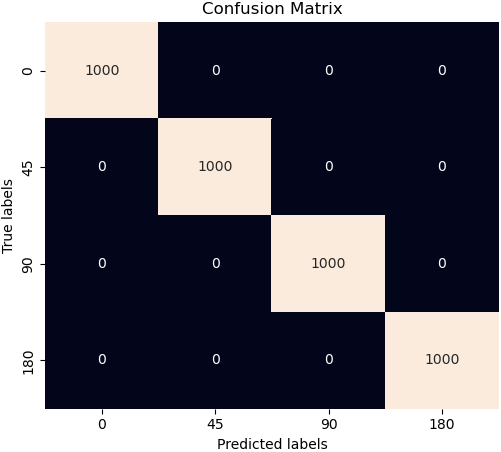}
			\caption{Distance 20ft}
		\end{subfigure}
		\begin{subfigure}[h]{0.32\linewidth}
			\includegraphics[width=\linewidth]{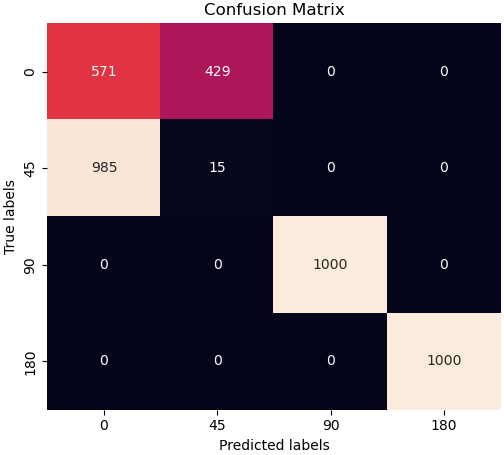}
			\caption{Distance 30ft}
		\end{subfigure}
		\begin{subfigure}[h]{0.32\linewidth}
			\includegraphics[width=\linewidth]{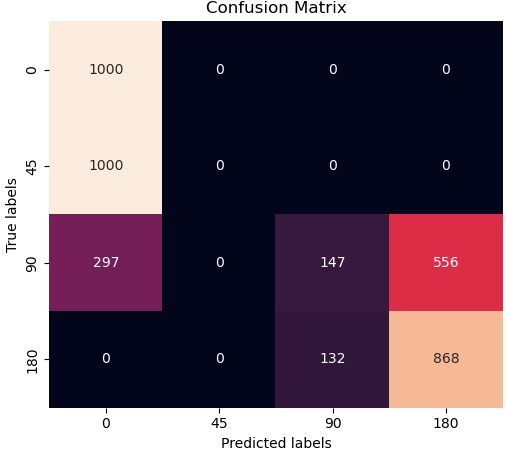}
			\caption{Distance 40ft}
		\end{subfigure}
		\begin{subfigure}[h]{0.32\linewidth}
			\includegraphics[width=\linewidth]{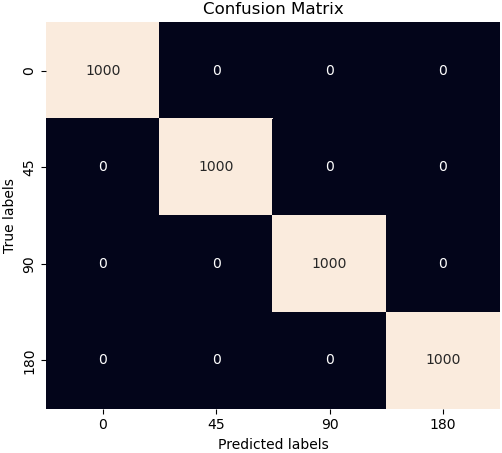}
			\caption{Distance 50ft}
		\end{subfigure}
		\caption{Predicting transmission angle}
		\label{3.7}
	\end{figure}

	Similarly, we evaluated the model to predict the angle of transmission. We needed to evaluate each distance group separately as `$10$ feet, $45$\textdegree' and `$20$ feet, $45$\textdegree' has very different data signatures. For $10$, $20$, and $50$ feets the model was very accurate in classifying the transmission angle. However, the model significantly misclassifies for $30$ and $40$ feet between angles of transmissions. Figures \ref{3.7}a-e detail the results regarding the classification of the transmission angle.

\begin{figure}
		\begin{subfigure}[h]{0.49\linewidth}
			\includegraphics[width=\linewidth]{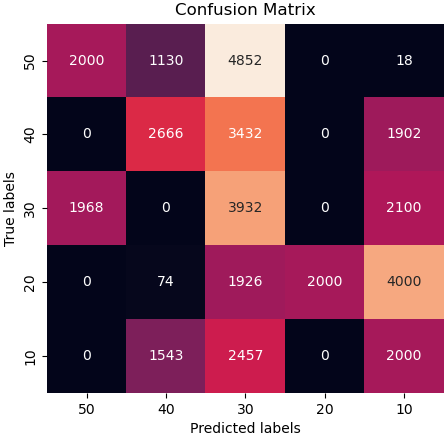}
			\caption{Distance prediction}
		\end{subfigure}
		\hfill
		\begin{subfigure}[h]{0.49\linewidth}
			\includegraphics[width=\linewidth]{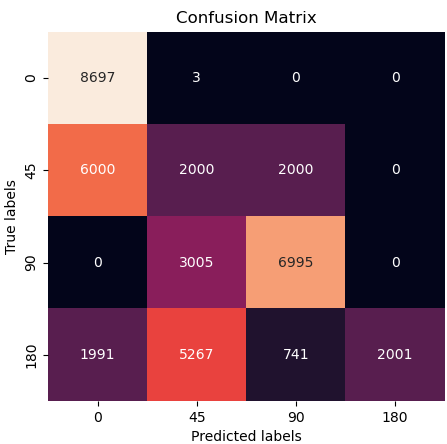}
			\caption{Angle prediction}
		\end{subfigure}
		\caption{Predicting transmission distance and angle at once}\label{3.8}
	\end{figure}

	In the same paradigm, we train the model with all 20 classes of data. We label each sample with two labels one for distance and another for angle. The target of the model is to uniquely identify angles and distances at once. figure \ref{3.8}a and b show that the results are much poorer. Notably, this is a much harder task. Here the model is expected to put `$10$ feet, $0$\textdegree', `$10$ feet, $45$\textdegree', `$10$ feet, $90$\textdegree', and `$10$ feet, $180$\textdegree' in the same $10$ feet class. Naturally, the same volume of training data is not sufficient as the simpler distance-only case. From observing trends of training with too little data (Figure \ref{3.6}b), we can generalize that with larger training data this performance is expected to improve.

	\subsection{Offline classification}
\begin{figure}
		\begin{subfigure}[h]{0.99\linewidth}
			\includegraphics[width=\linewidth]{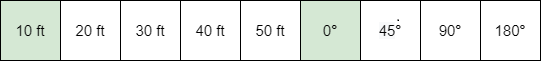}
			\caption{Multiclass classification}
		\end{subfigure}
		\hfill
		\begin{subfigure}[h]{0.99\linewidth}
			\includegraphics[width=\linewidth]{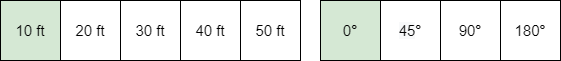}
			\caption{Multihead classification}
		\end{subfigure}
		\caption{The difference between the two classification tasks}\label{3.9}
	\end{figure}

	In this section, we perform offline classification of the saved data. When we can iterate over the data multiple times, the same volume of data yields more insight. Thus, when possible it's more advantageous to utilize offline classification. The transmission data is sequential and has a temporal component that can be exploited. Leveraging this fact, we trained two LSTM \cite{hochreiter1997long} based architecture to classify the hardest task, distance, and angle identification at once. The problem is a multi-label classification task. Each sample has two labels, distance, and angle. The first model was a multiclass class model that was trained to identify two labels from nine possible labels (figure \ref{3.9}a), whereas the second model was a multi-head model that was trained to perform two separate classification tasks at once (figure \ref{3.9}a).  
	
	The multiclass model was much better than the online kitsune model but not quite as well as the multi-head model for the PCA and the empirically selected features. It performs exceptionally well with the MRMR features.
	The multi-head model overall provided better results. We also observed that The features selected through MRMR performed better during training but performed poorer than the empirically selected features during testing. The features extracted through PCA however present the best results. The multiclass model with the MRMR features and the multi-head model with the PCR features perform comparably and are significantly better than the other approaches.  Table \ref{table1} details the overall accuracy of the different approaches.

	\begin{table}[!htb]
		\centering
		% \resizebox{\linewidth}{!}{
			\begin{tabular}{| l l l l|}
				\hline
				Model               & Features               & Distance                 & Angle                    \\ \hline
				\textit{Kitsune}    & Empirical              & 31.5\%                   & 49.2\%                   \\
				                    & MRMR                   & 43.0\%                   & 56.9\%                   \\
				                    & PCR                    & 33.5\%                   & 59.0\%                   \\
				\textit{Multiclass} & Empirical              & 70.9\%                   & 88.0\%                   \\
				                    & \textcolor{blue}{MRMR} & 97.8\%                   & \textcolor{blue}{99.0\%} \\
				                    & PCR                    & 80.9\%                   & 87.1\%                   \\
				\textit{Multihead}  & Empirical              & 93.0\%                   & 98.6\%                   \\
				                    & MRMR                   & 88.5\%                   & 92.4\%                   \\
				                    & \textcolor{blue}{PCR}  & \textcolor{blue}{98.7\%} & 98.9\%                   \\ \hline
			\end{tabular}
			% 	}
		\caption{ Classification accuracy for the different approaches} \label{table1}
	\end{table}
	
\begin{figure}
		\begin{subfigure}[!htb]{0.49\linewidth}
			\includegraphics[width=\linewidth]{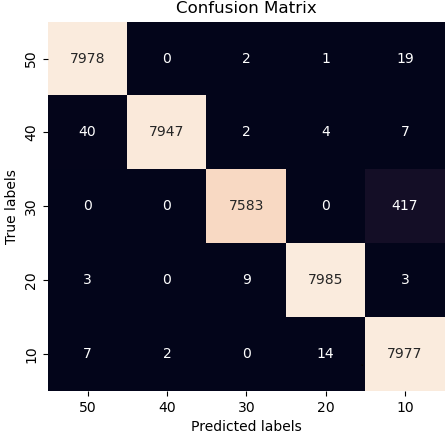}
			\caption{Distance prediction}
		\end{subfigure}
		\hfill
		\begin{subfigure}[h]{0.49\linewidth}
			\includegraphics[width=\linewidth]{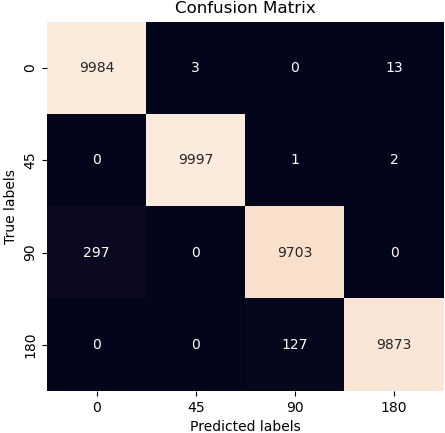}
			\caption{Angle prediction}
		\end{subfigure}
		\caption{Predicting transmission distance and angle with the multi-head model}\label{3.10}
	\end{figure}

	Figure \ref{3.10} shows the confusion matrix for the multi-head classification with the PCA features. 
	
	\section{Conclusion}
	Various 5G use cases, such as virtual reality, real-time distributed UHD gaming, and tactile internet for remote surgery, need higher bandwidth and less latency in a highly mobile environment. Providing the requirements from these 5G use cases needs fiber-like connectivity. To this end, Fixed Broadband Wireless Access in mmWave Band seems a very promising approach. In this paper, we analyzed the performance of mmWave-based devices operating in the GHz frequency band (i.e., from $57$-$71$ GHz in the USA). We gathered the statistical data and used machine learning approaches to classify them. While analyzing the data features for the classification, we observed that the features for transmission at $45$\textdegree  and $90$\textdegree angles diverge a lot more than transmission at $180$\textdegree angles.  We collected sample data for transmission distances of $25$ and $35$ feet with the goal to train a regression on the available data and predict the distance and angle correctly without training explicitly on the data classes. Our reported models could not be generalized to achieve this goal. This is something that we wish to address in future research.
	Building on machine learning approaches the radio can identify and self-configure to achieve better performance in an adverse transmission configuration. This phenomenon is worth investigating in future research.

	\bibliographystyle{IEEEtran}
	\bibliography{ref}
	
\end{document}